\documentstyle[aps,twocolumn]{revtex}

\newcommand{\be}{\begin{equation}}
\newcommand{\ee}{\end{equation}}
\newcommand{\bea}{\begin{eqnarray}}
\newcommand{\eea}{\end{eqnarray}}

\newcommand{\mtwo}[4] {\left( \matrix{{#1}&{#2}\cr{#3}&{#4}} \right)}
\newcommand{\mthree}[9]
  {\left( \matrix{{#1}&{#2}&{#3}\cr{#4}&{#5}&{#6}\cr{#7}&{#8}&{#9}} \right)}

\begin{document}

\wideabs{

\title{Universal simulation of Markovian quantum dynamics}

\author{
Dave Bacon,$\!^{1,2}$ Andrew M.~Childs,$^{3,4}$
Isaac L.~Chuang,$^4$ Julia Kempe,$\!^{1,5,6}$
Debbie W.~Leung,$^{4,7,8}$ and Xinlan Zhou$^{4,9}$
}

\address{
Departments of Chemistry,$^1$ Physics,$^2$ and Mathematics,$^5$ University
of California, Berkeley, CA 94720
\\
Physics Department, California Institute of Technology,$^3$ Pasadena, CA
91125
\\
IBM Almaden Research Center,$^4$ 650 Harry Road, San Jose, CA 95120
\\
\'Ecole Nationale Superieure des T\'el\'ecommunications,$^6$ Paris, France
\\
Quantum Entanglement Project, ICORP, JST, Edward Ginzton Laboratory$^7$
\\
and Applied Physics Department,$^9$ Stanford University, Stanford, CA 94305
\\
IBM T. J. Watson Research Center,$^8$ Yorktown Heights, NY 10598
}

\date{18 June 2001}

\maketitle


\begin{abstract}
Although the conditions for performing arbitrary unitary operations to
simulate the dynamics of a closed quantum system are well understood, the
same is not true of the more general class of quantum operations (also known
as superoperators) corresponding to the dynamics of open quantum systems.
We propose a framework for the generation of Markovian quantum dynamics and
study the resources needed for universality.  For the case of a single
qubit, we show that a single nonunitary process is necessary and sufficient
to generate all unital Markovian quantum dynamics, whereas a set of
processes parametrized by one continuous parameter is needed in general.  We
also obtain preliminary results for the unital case in higher dimensions.
\end{abstract}



} 

\section{Introduction}

The idea of simulating one quantum system with another was first suggested
in the early 1980's by Manin~\cite{Manin} and Feynman~\cite{Feynman}.  A
universal quantum computer can perform such simulation because of its
ability to apply arbitrary unitary transformations to arbitrary quantum
states.  The only necessary resources are single qubit gates and the
controlled-{\sc not} two-qubit gate~\cite{DiVincenzo}.  In fact, the
controlled-{\sc not} may be replaced by nearly any two-qubit
interaction~\cite{AnyTwo}, and the single-qubit gates can be reduced to a
finite set~\cite{Shor}.  Finiteness of the gate set is desirable because it
reduces the necessary computational resources and simplifies the
construction of fault tolerant gates.

Using a universal gate set, a quantum computer may simulate the time
sequence of operations corresponding to any unitary dynamics.  Such
simulation is provably efficient~\cite{Lloyd} and has been implemented in
the context of NMR quantum computation~\cite{Somaroo}.

However, quantum systems may undergo interesting processes which are not
unitary due to interactions with their environments.  Evolution of such {\em
open quantum systems} is described by {\em quantum operations} (or
superoperators).  Understanding such dynamics is important for studying
quantum noise processes~\cite{Gardiner}, designing quantum error correcting
codes~\cite{ErrorCor}, and performing simulations of open quantum systems,
such as of thermal equilibration~\cite{Terhal}.

Clearly, creation of arbitrary quantum operations and simulation of
arbitrary quantum dynamics using a simple set of primitives are desirable
goals.  However, it is more difficult to describe a notion of universality
for general quantum operations than for unitary operations alone.  Unlike
unitary operations, which form a Lie group, quantum operations comprise a
semigroup due to their irreversibility.  The lack of inverse operations for
semigroups is troublesome, and it is less obvious how best to combine
quantum operations to form new ones.

A simple recipe for implementing a general quantum operation follows
from its unitary representation: any quantum operation may be written
as a unitary operation on an extended system with a trace over the
extra degrees of freedom.  As is well known, this procedure only
requires an ancillary system of dimension equal to the square of the
dimension of the system of interest to produce arbitrary quantum
operations.
However, our goal is to consider as a resource a small subset of
nonunitary quantum dynamics applied to the system only, without the
need to control the extra degrees of freedom.  Such restrictions are
important in many applications, including the experimental simulation
of quantum systems.  For this reason, we exclude the technique arising
from the unitary representation when building arbitrary nonunitary
quantum dynamics.

In this paper, we begin to study possible methods for simulating the
dynamics of open quantum systems by some time sequence of operations.  We
take the approach of considering only processes which result from
interaction with a Markovian environment in the Born approximation.  We
refer to this class of dynamics as {\em Markovian quantum dynamics} and
refer to the semigroup they comprise as a {\em Markovian semigroup}.  Such
processes have a convenient description in terms of their {\em generators},
a concept analogous to the Hamiltonian of unitary dynamics.  Therefore,
simulation of Markovian quantum dynamics is reduced to building generators
for Markovian semigroups.

We define two allowed procedures for transforming semigroup generators, {\em
linear-combination} and {\em unitary conjugation}.  Using these procedures, we
show how to build more complicated generators from simple ones, and we
explore in detail the required resources for the case of a single quantum
bit.

The structure of the paper is as follows: in Section~\ref{sec:qops} we
present representations of quantum operations and the description of
Markovian quantum dynamics in terms of semigroups.  Then, in
Section~\ref{sec:framework}, we describe the procedures of linear-combination
and unitary conjugation.  Using these procedures, we present universal sets
of generators for unital Markovian quantum dynamics on a qubit in
Section~\ref{sec:unital} and for general Markovian quantum dynamics on a
qubit in Section~\ref{sec:general}.  Finally, we conclude with some open
questions and directions for further investigation.

\section{Quantum operations and Markovian semigroups}
\label{sec:qops}

A quantum state is described by a density matrix $\rho$ which is
positive semidefinite and has ${\rm tr} \rho = 1$.
The most general state change of a quantum system, a quantum operation, is a
linear map $\cal E$ which is trace preserving and completely positive.
${\cal E}$ acts on $\rho$ to produce a state ${\cal E}(\rho)$.
There are many representations for such a map.  The operator sum
representation
\be
  {\cal E}(\rho) = A_k \rho A_k^\dag
\label{eq:osr}
\ee
(note that we use Einstein summation where appropriate) --- and its
corresponding fixed-basis form~\cite{Chuang} --- is convenient in that the
constraints of trace preservation and complete positivity may be simply
expressed.  For example, complete positivity is inherent in
Eq.~(\ref{eq:osr}) and trace preservation is equivalent to $A_k^\dagger A_k
= I$, where $I$ is the identity matrix.  However, the composition of two
operator sum representations is complicated, usually resulting in a rapidly
increasing number of terms.  On the other hand, a manifestly linear
representation
\be
  ({\cal E}(\rho))_{ab} = M_{(ab)(cd)} \rho_{cd}
\label{eq:qoplin}
\ee
(where $M$ is a matrix with composite indices) makes composition of
operations trivial, yet obfuscates the constraints~\cite{CP}.

Instead of considering all possible dynamics, we will simplify the problem
by focusing on Markovian quantum dynamics.  We describe these processes
informally here, saving a more complete presentation based
on~\cite{Davies76,Davies80,Alicki} for Appendix~\ref{app:qops}.  Every such
process corresponds to some interaction which, if applied for a duration
$t$, induces a quantum operation ${\cal E}_t$.  The class of quantum
operations ${\cal E}_t$ forms a Markovian semigroup.  The time $t$ may vary
continuously.  The operations must be stationary and Markovian, such that
\be
  {\cal E}_s {\cal E}_t = {\cal E}_{s+t}
\,.
\label{eq:markov}
\ee
Here ${\cal E}_s {\cal E}_t$ denotes composition of the operations, i.e.,
${\cal E}_s \circ {\cal E}_t$.  Each Markovian semigroup describes the
dynamics resulting from some interaction with a Markovian environment in the
Born approximation.

Note that this terminology differs slightly from that used elsewhere.  For
example, Davies does not include the constraint of trace preservation when
defining a Markov semigroup in~\cite{Davies76} and, curiously, uses
``Markov'' to mean ``unital'' in~\cite{Davies80}.  For a precise definition
of Markovian semigroups as used in this paper, see Appendix~\ref{app:qops}.

The advantage of considering only Markovian semigroups is that they are
uniquely determined by their generators.  The generator $\cal Z$ of ${\cal
E}_t$ is defined by its action on an arbitrary input $\rho$,
\be
  {\cal Z}(\rho)=\lim_{t \downarrow 0} {{\cal E}_t(\rho) - \rho \over t}
\,.
\label{eq:generator}
\ee
In a sense, $\cal Z$ can be thought of as the ``Hamiltonian'' corresponding
to ${\cal E}_t$.  Exponentiation gives
\be
  {\cal E}_t=e^{{\cal Z} t} \equiv \lim_{n \to \infty}
             \left( {\cal I} - {t \over n} {\cal Z} \right)^{-n}
\,,
\label{eq:exponent}
\ee
where $\cal I$ is the identity quantum operation.  The generator also
satisfies the differential equation
\be
        {\partial \rho(t) \over \partial t} = {\cal Z}(\rho(t))
\,,
\label{eq:diffeq}
\ee
which is known as a {\em master equation}.  Through this analysis,
simulating ${\cal E}_t~\forall t \ge 0$ is reduced to simulating its
generator.

Gorini, Kossakowski, and Sudarshan have shown that $\cal Z$ is the
generator of a Markovian semigroup on an $N$ dimensional Hilbert
space if and only if it can be written in the form~\cite{GKS}
\be
  {\cal Z}(\rho) = -i [H,\rho]
                   + a_{\alpha \beta}
                     ([F_\alpha \rho, F_\beta^\dagger]
                     +[F_\alpha, \rho F_\beta^\dagger])
\,,
\label{eq:gks}
\ee
where $a_{\alpha \beta}$ is an $(N^2-1) \times (N^2-1)$ positive matrix
(with $\alpha,\beta \in [1,N^2-1]$) and $\{F_\alpha\}$ is a linear basis of
traceless operators on the density matrices.  We refer to the matrix
$a_{\alpha\beta}$ as the {\em GKS matrix}.  For related formulations, such
as the ``diagonal'' form introduced by Lindblad (which also applies to
countably infinite-dimensional systems), see~\cite{Alicki,Lindblad}.
Physically, $H$ corresponds to unitary dynamics which can be produced by a
system Hamiltonian as well as unitary dynamics induced by a coupling between
the system and the bath --- the so-called Lamb shift.

It will greatly simplify the discussion to choose a Hermitian basis which is
orthonormal under the trace norm.  Such a basis is assumed for the rest of
the paper.  Therefore,
\be
  {\rm tr}(F_\alpha F_\beta^\dag) = \delta_{\alpha\beta}
\,,
\label{eq:trace}
\ee
and ${\rm tr}(F_\alpha) = 0$.  Note that we can always reduce a GKS matrix
which is expressed in an overcomplete or non-orthonormal traceless basis to
a representation involving a linearly independent orthonormal traceless
basis.

There are other ways to describe the generator of a Markovian
semigroup.  For example, ${\cal Z}(\rho)$ may always be written as an
affine transformation of $\rho$, just as any quantum operation can be
written as a linear transformation as in Eq.~(\ref{eq:qoplin}).  In
this paper, we find it simplest to represent generators by the GKS
matrix, and we describe the relationship between the GKS matrix and
the affine representation in Appendix \ref{app:gks2lin}.

To make our description of Markovian quantum dynamics concrete, we present
some important examples of qubit noise processes~\cite{QubitOps}.  We choose
the basis $\{F_\alpha\}$ to be the normalized Pauli operators ${1 \over
\sqrt{2}} \{\sigma_x, \sigma_y, \sigma_z\}$, and we write the density matrix
of a qubit as
\be
\rho = \mtwo{\rho_{00}}{\rho_{01}}{\rho_{10}}{\rho_{11}}
\,.
\ee
The first process, {\em phase damping}, acts on a qubit as
\be
  {\cal E}^{\rm PD}_t(\rho)
     = \mtwo{\rho_{00}}{e^{-\gamma t} \rho_{01}}
            {e^{-\gamma t} \rho_{10}}{\rho_{11}}
\,,
\ee
where $\gamma$ is a decay constant and $t$ is the duration of the process.
The generator has a GKS matrix with $a_{33}^{\rm PD} = {\gamma \over 2}$ and
all other $a_{\alpha \beta}^{\rm PD} = 0$.  The second example is the {\em
depolarizing channel}, which acts on a qubit as
\be
  {\cal E}^{\rm DEP}_t(\rho) =
    \mtwo{1+e^{-\tilde\gamma t}(\rho_{00}-\rho_{11}) \over 2}
         {e^{-\tilde\gamma t}\rho_{01}}
         {e^{-\tilde\gamma t}\rho_{10}}
         {1+e^{-\tilde\gamma t}(\rho_{11}-\rho_{00}) \over 2}
\,.
\ee
Its GKS matrix has the nonzero elements $a_{11}^{\rm DEP}=a_{22}^{\rm
DEP}=a_{33}^{\rm DEP}=\tilde\gamma/4$.  Our final example is {\em amplitude
damping}, which acts on a qubit as
\be
  {\cal E}^{\rm AD}_t(\rho) =
    \mtwo{\rho_{00}+(1-e^{-\Gamma t}) \rho_{11}}
         {e^{-\Gamma t/2} \rho_{01}}
         {e^{-\Gamma t/2} \rho_{10}}
         {e^{-\Gamma t} \rho_{11}}
\,.
\label{eq:ad}
\ee
The GKS matrix $a_{\alpha \beta}^{\rm AD}$ is given by
\be
     {\Gamma \over 4} \mthree{1}{-i}{0}
                             {i}{1}{0}
                             {0}{0}{0}
\,.
\ee
Note that the GKS matrix is real and diagonal for phase damping and the
depolarizing channel and has rank one for phase damping and amplitude
damping.

\section{Composition framework: linear-combination and unitary conjugation}
\label{sec:framework}

Recall that our goal is to find a simple way of combining as few primitive
${\cal E}^i$ as possible to produce all possible $\cal E$ via some time
sequence of operations.  To make this problem well-posed, we must choose
reasonable methods for composing quantum operations to make new ones.  We
have not found a simple way to express the composition of two semigroup
processes of finite duration, and such composition need not preserve
Markovity.  However, a natural way to combine semigroup processes is by a
procedure we call {\em linear-combination}: the processes act one after another
for small amounts of time.  In the limit of infinitesimal time steps, two
processes ${\cal E}^a_t$ and ${\cal E}^b_t$ can be combined to produce
\be
  {\cal E}^{a+b}_t \equiv \lim_{n \to \infty} ({\cal E}^a_{t/n}
                                               {\cal E}^b_{t/n})^n
\,,
\ee
where ${\cal E}^{a+b}_t$ forms a Markovian semigroup if ${\cal E}^{a}_t$ and
${\cal E}^{b}_t$ do.  Moreover, if ${\cal E}^{a}_t$ and ${\cal E}^{b}_t$
have generators $\cal A$ and $\cal B$, then applying Lie's
product formula to the generators,
\be
  \lim_{n \to \infty} \left( e^{{\cal A} t/n} e^{{\cal B} t/n} \right)^n
  = e^{({\cal A} + {\cal B}) t}
\label{eq:trotter}
\,.
\ee
In other words, the generator of a process formed by linear-combination is
the sum of the constituent generators.  The generalization to produce a
positive sum of any finite number of generators is straightforward.  When
all generators are expressed in the form of Eq.~(\ref{eq:gks}) using the
same basis $\{F_\alpha\}$ (as we assume for the rest of the paper),
linear-combination corresponds to a positive sum of the GKS matrices of the
constituent generators.

We also assume the capability to apply arbitrary unitary operations to the
system, since these tasks are feasible and well understood.  Using
linear-combination, we may produce the two terms in Eq.~(\ref{eq:gks})
separately.  Assuming the ability to create any unitary dynamics, it remains
to generate the second term under the assumption $H=0$.

We now turn to the second procedure to transform the GKS matrix,
called {\em unitary conjugation}.  This procedure transforms $\cal E$
according to
\be
  {\cal U}^\dag {\cal E} {\cal U}
\,,
\label{eq:unitary}
\ee
where ${\cal U}(\rho) = U \rho U^\dag$ for some unitary operator $U$.  Note
that unitary conjugation preserves all the Markovian semigroup properties.
We will see that the effect of unitary conjugation is to apply $\cal E$ in a
different basis, producing a new operation which may be used on its own or
in linear-combination.  To understand how the GKS matrix transforms, we prove
the following theorem:
\begin{quote}
{\bf Theorem 1}~{\em
For an $N$ dimensional system, unitary conjugation by $U \in SU(N)$ results
in conjugation of the GKS matrix by a corresponding element in the adjoint
representation of $SU(N)$.}
\end{quote}

{\em Proof:} Suppose the Markovian semigroup has generator $\cal A$ and GKS
matrix $a_{\alpha\beta}$.  Conjugation by $U$ results in the evolution
\bea
{\cal U}^\dag e^{{\cal A} t} {\cal U}
  &=& \lim_{n \to \infty} {\cal U}^\dagger
      \left({\cal I} - {t \over n} {\cal A} \right)^{-n} {\cal U} \\
  &=& \lim_{n \rightarrow \infty}
      \left({\cal I} - {t \over n} {\cal U}^\dagger {\cal A} {\cal U}
     \right)^{-n}
\,.
\eea
In other words, the new generator is ${\cal A}'={\cal U}^\dag {\cal A} {\cal
U}$.  Expressed in the form of Eq.~(\ref{eq:gks}) (with $H=0$), we find
\bea
  {\cal A}'(\rho) = a_{\alpha \beta}
    (&&[U^\dag F_\alpha U \rho,
        U^\dag F_\beta^\dag U] \nonumber \\
    +&&[U^\dag F_\alpha U, \rho
        U^\dag F_\beta^\dag U])
\label{eq:aprime}
\,.
\eea
Evidently this unitary conjugation induces a change of basis $F_\alpha \to
U^\dag F_\alpha U$, which is still Hermitian, orthonormal, and traceless.
We can expand the new basis in terms of the old one:
\be
  U^\dag F_\alpha U = c_{\alpha \gamma} F_\gamma
\,.
\label{eq:bexpand}
\ee
This implies
\be
  U^\dag F_\alpha F_\beta^\dag U =
    c_{\alpha \gamma} c_{\beta \nu}^* F_\gamma F_\nu^\dag
\label{eq:uffu}
\,.
\ee
Taking the trace of Eq.~(\ref{eq:uffu}), and using the orthonormality of the
$F_\alpha$ (Eq.~(\ref{eq:trace})),
\be
  c_{\alpha \gamma } c_{\beta \gamma}^* = \delta_{\alpha \beta}.
\ee
In other words, $c_{\alpha \gamma}$ is a unitary matrix.  Further, by
substituting Eq.~(\ref{eq:bexpand}) into Eq.~(\ref{eq:aprime}), we obtain
the transformed GKS matrix
\be
  a_{\gamma \nu}' = c_{\alpha \gamma} a_{\alpha \beta} c_{\beta \nu}^*
\,.
\ee
Denoting the matrices $a_{\alpha \beta}$ and $c_{\alpha \beta}$ by $A$ and
$C$, $A' = C^T A C^*$.  The effect of unitary conjugation is to conjugate
the original GKS matrix by $C^T$.

Note that $C$ is not arbitrary, but is determined by $U$ in the following
manner.  Suppose we choose $\{F_\alpha\}$ to be the generators of $SU(N)$.
Then we have
\be
  [F_\alpha, F_\beta] = i f_{\alpha\beta\gamma} F_\gamma
\,,
\label{eq:lie}
\ee
where $f_{\alpha\beta\gamma}$ are the real structure constants for the Lie
algebra generated by $\{F_\alpha\}$.  Setting $U=e^{ir_\gamma F_\gamma}$, we
find to first order in an infinitesimal $r_\gamma$ that
\bea
  U^\dag F_\alpha U
    &=& (I-ir_\gamma F_\gamma) F_\alpha (I+ir_\gamma F_\gamma) \nonumber\\
    &=& F_\alpha - i r_\gamma [F_\gamma, F_\alpha] \nonumber\\
    &=& F_\alpha - i r_\gamma (i f_{\gamma\alpha\beta} F_\beta)
\,.
\eea
Thus
\bea
  (C^T)_{\alpha\beta}
    = \delta_{\alpha\beta} + i r_\gamma (i f_{\gamma\alpha\beta})
\eea
is in a Lie group generated by the matrices $(G_\gamma)_{\alpha\beta} = i
f_{\gamma\alpha\beta}$.

It is an elementary fact of group theory that these $G_\gamma$ are the
generators of the adjoint representation of the $F_\alpha$ algebra.
Therefore, $U=e^{ir_\gamma F_\gamma} \in SU(N)$ induces conjugation of the
GKS matrix by $C^T = e^{ir_\gamma G_\gamma}$ in the adjoint representation
of $SU(N)$. \hfill~$\Box$

As an example of an application of these two methods, we can perform
linear-combination of amplitude damping, ${\cal E}^{\rm AD}$
(Eq.~(\ref{eq:ad})), and damping in the opposite direction, ${\cal X}^\dag
{\cal E}^{\rm AD} {\cal X}$, where ${\cal X}(\rho) = \sigma_x \rho
\sigma_x$, to simulate generalized amplitude damping with an arbitrary
mixture of the ground and excited states as the fixed point.

\section{Unital Markovian quantum dynamics}
\label{sec:unital}

We now use the resources we have defined to simulate Markovian quantum
dynamics.  We first consider {\em unital} processes, those that fix the
identity.  Well-known unital processes on a qubit include phase damping and
the depolarizing channel.

We first characterize unitality for GKS matrices.  ${\cal Z}$ is the
generator of a unital Markovian semigroup iff ${\cal Z}(I)=0$.  Using
Eq.~(\ref{eq:gks}) (with $H=0$), Eq.~(\ref{eq:lie}), and the antisymmetry of
$f_{\alpha\beta\gamma}$, we find
\bea
  {\cal Z}(I)
    &=& 2 a_{\alpha\beta} [F_\alpha, F_\beta^\dag]
     =  2 a_{\alpha\beta} [F_\alpha, F_\beta] \nonumber \\
    &=& 2 i a_{\alpha\beta} f_{\alpha\beta\gamma} F_\gamma
     =  2 i \sum_{\alpha < \beta} (a_{\alpha\beta}-a_{\beta\alpha})
          f_{\alpha\beta\gamma} F_\gamma \nonumber \\
    &=& -4 \sum_{\alpha < \beta} {\rm Im}(a_{\alpha\beta})
          f_{\alpha\beta\gamma} F_\gamma
\,.
\eea
By orthonormality of the $F_\alpha$, ${\cal Z}(I)=0$ iff
\be
  \sum_{\alpha < \beta} {\rm Im}(a_{\alpha\beta}) f_{\alpha\beta\gamma} = 0
  \quad\forall\gamma
\,.
\label{eq:nscond}
\ee
In general, reality of the GKS matrix is a sufficient condition for
unitality.  When $N=2$, the sum has only one term, so that this condition is
also necessary.  Thus the unital Markovian semigroups on a qubit are exactly
those generated by real GKS matrices.

For $N>2$, reality of the GKS matrix is not necessary for the corresponding
process to be unital.  In Appendix~\ref{app:unital}, we give an example of a
unital operation of dimension $N=3$ for which the GKS matrix is not real.
Since $SU(N)$ for $N \ge 3$ contains an isomorphic copy of $SU(3)$, it
follows that there are unital Markovian semigroups generated by complex GKS
matrices for all $N \ge 3$.  The set of Markovian semigroups generated by
real GKS matrices is a {\em proper} subset of the set of all unital
Markovian semigroups.

We now focus on Markovian quantum dynamics on a qubit and consider the
effects of unitary conjugation.  We take $F_{1,2,3} = \sigma_{x,y,z}$ as
before.  The generators are represented by real positive semidefinite  
GKS matrices $A$.  The transformations induced by unitary conjugation 
are simply
\be
  A' =  e^{i r_\gamma G_\gamma} \, A \, e^{-i r_\gamma G_\gamma}
\,,
\ee
where
\bea
\label{eq:SO(3)}
  G_1 &=& i\mthree{0}{0}{0}
                  {0}{0}{1}
                  {0}{-1}{0}
\nonumber \\
  G_2 &=& i\mthree{0}{0}{-1}
                  {0}{0}{0}
                  {1}{0}{0}
\nonumber \\
  G_3 &=& i\mthree{0}{1}{0}
                  {-1}{0}{0}
                  {0}{0}{0}
\eea
can be found using $f_{\alpha\beta\gamma} = \epsilon_{\alpha\beta\gamma}$
for the Pauli matrices.  Note that the $G_\gamma$'s are simply the
generators of $SO(3)$ --- as is well known, $SO(3)$ is the adjoint
representation of $SU(2)$.

Having characterized unital Markovian quantum dynamics on a qubit, we are
ready to state the following:
\begin{quote}
{\bf Theorem 2}~{\em
All unital Markovian quantum dynamics on a single qubit can be simulated
with linear-combination and unitary conjugation given phase damping as a
primitive operation.}
\end{quote}

{\em Proof:}
When $A$ is positive semidefinite and real, $A = O D O^T$ for some 
$O \in SO(3)$ and $D$ diagonal with diagonal entries $d_i \geq 0$. 
Let $A^{\rm PD}$ denote the GKS matrix for phase damping.  Then
\bea
  A = {2 \over \gamma} (
      && d_1 O e^{i\frac{\pi}{2} G_2} A^{\rm PD} e^{-i\frac{\pi}{2} G_2} O^T
      \nonumber\\
     +&& d_2 O e^{i\frac{\pi}{2} G_1} A^{\rm PD} e^{-i\frac{\pi}{2} G_1} O^T
     +   d_3 O A^{\rm PD} O^T )
\,.
\eea
Thus the GKS matrix for any unital Markovian semigroup can be simulated
using linear-combination of the unitary conjugates of phase damping.
\hfill~$\Box$

Note that any unital Markovian semigroup with rank one GKS matrix can
be used in place of phase damping.

\section{General Markovian quantum dynamics on a qubit}
\label{sec:general}

We now consider general Markovian quantum dynamics on a single qubit.  We
prove the following:
\begin{quote}
{\bf Theorem 3}~{\em
To simulate arbitrary Markovian quantum dynamics on a qubit with
linear-combination and conjugation by unitaries, it is necessary and sufficient
to be able to perform all operations with GKS matrix $A(\theta) = \vec
a(\theta) \, \vec a(\theta)^\dagger$, where $\vec a(\theta) \equiv
(\cos\theta,i\sin\theta,0)^T$ and $\theta \in [0,{\pi \over 4}]$.}
\end{quote}

{\em Sufficiency:} Let $A$ be the GKS matrix to be simulated.  Since
$A$ is positive semidefinite, we apply the spectral decomposition to 
express $A$ as a positive sum of outer products,
\be
  A = \sum_k \lambda_k \vec a_k \vec a_k^\dag
\,,
\ee
where $\lambda_k \ge 0$ and $\vec a$'s are normalized to unit length.
By linear-combination, it suffices to simulate all $\vec a \vec
a^\dag$.  Separating into real and imaginary parts, we may write any
such vector as
\be
  \vec a=\vec a^R+i\vec a^I
\,.
\ee

Because $\vec a$ only appears in outer products, the overall phase of $\vec
a$ is irrelevant.  In other words, we have $\vec a' \vec a'^\dagger = \vec a
\vec a^\dagger$, where
\bea
  \vec a' = e^{i\psi} \vec a =  &&(\vec a^R \cos\psi - \vec a^I \sin\psi)
  \nonumber \\
                              +i&&(\vec a^R \sin\psi + \vec a^I \cos\psi)
\,.
\eea
This transformation maps the two parameters
\bea
  k_1 &\equiv& |\vec{a}^R|^2-|\vec{a}^I|^2 \\
  k_2 &\equiv& 2 \vec{a}^{R\dag} \vec{a}^I
\eea
according to
\be
  \left(\matrix{k_1' \cr k_2'}\right)
  = \mtwo{\cos 2\psi}{-\sin 2\psi}{\sin 2\psi}{\cos 2\psi}
    \left(\matrix{k_1 \cr k_2}\right)
\label{eq:phasetrsf}
\,.
\ee
Because we may choose $\psi$ arbitrarily, we make the choice
\be
\tan 2\psi = -k_2/k_1
\,,
\label{eq:orthcond}
\ee
such that $k_2'= 0$, in which case $\vec a'^R$ and $\vec a'^I$ are
orthogonal.  Moreover, we can choose $k_1' = k_1/\cos2\psi \geq 0$ such that
$|\vec{a}^{'R}| \geq |\vec{a}^{'I}|$.  Thus, without loss of generality, we
may assume that $\vec a$ has a real part no shorter than its imaginary part,
and that the two parts are orthogonal.

Performing a unitary transformation on the operation effects conjugation
by $G \in SO(3)$:
\be
  \vec a \vec a^\dag \to G \vec a \vec a^\dag G^T
  = (G \vec a)(G \vec a)^\dag
\,.
\ee
Because $G$ is real, it does not mix the real and imaginary parts of $\vec
a$ in $G \vec a$.  In other words, $G$ simultaneously rotates the two
vectors $\vec a^R$ and $\vec a^I$.  Therefore, $G$ can always be chosen to
align $\vec a^R$ with the $+x$ axis and $\vec a^I$ with the $+y$ axis, so
that we can write
\be
\label{eq:aform}
  G \vec a
    = \vec a(\theta) \nonumber
    = (\cos\theta, i\sin\theta, 0)^T
\,,
\ee
where $\theta \in [0,{\pi \over 4}]$.

{\em Necessity:}  The $A(\theta)$, being rank one, are extreme in the convex
cone of all the positive matrices.  Thus linear-combination cannot be used to
compose a new $A(\theta')$.

Because scalar multiplication of $\vec a$ by a phase commutes with an
$SO(3)$ transformation, it suffices to show that given a phase $\psi$ and
rotation $G$ such that $G e^{i\psi} \vec a(\theta) = \vec a(\theta')$ with
$\theta, \theta' \in [0,{\pi \over 4}]$, then $\theta = \theta'$.  To see
this, note that given $k_2 = k_2' = 0$ and $k_1, k_1' \ge 0$ in
Eq.~(\ref{eq:phasetrsf}), the phase transformation must be trivial.  Thus
the real and imaginary parts are unchanged by $e^{i\psi}$, and have to be
unchanged by $G$ to remain aligned with the $+x$ and $+y$ axes.  Therefore,
$\theta'=\theta$. \hfill~$\Box$

Note that the set of required operations includes amplitude damping,
with $\vec a^{\rm AD} = \vec a(\pi/4) = (1,i,0)^T / \sqrt{2}$, and
phase damping about the $x$ axis, with $\vec a^{\rm PD'} = \vec a(0) =
(1,0,0)^T$.

We have seen that a one parameter set of generators is necessary and
sufficient to simulate all Markovian quantum dynamics on a qubit.  With this
result in mind, we can estimate the number of parameters in the universal
set of generators needed for an $N$ dimensional quantum system.  Generating
the $(N^2-1)\times(N^2-1)$ GKS matrix can be reduced to generating all those
with rank one by linear-combination.  Thus we have to obtain all (normalized)
complex $(N^2-1)$-vectors $\vec a$ up to a phase, with $2(N^2-1)-2$ free
parameters.  As there are $N^2-1$ degrees of freedom in $SU(N)$, unitary
conjugation will eliminate at most $N^2-1$ parameters, leaving $N^2-3$
parameters to be obtained, perhaps by a continuous set of primitives.  In
the case of a qubit ($N=2$), we have seen that this bound is tight.

\section{Conclusions and open questions}
\label{sec:conclude}

In the search for a way to simulate the dynamics of open quantum systems
using a simple set of primitives, we have set up a framework to study the
notion of universality for Markovian quantum dynamics.  We have shown how
the generators of Markovian semigroups transform under the composition
procedures of linear-combination (Eq.~(\ref{eq:trotter})) and unitary
conjugation (Theorem 1).  For the case of a single qubit, we
have shown that one primitive unital operation suffices to simulate all
other unital operations (Theorem 2), and we have exhibited a
necessary and sufficient single parameter universal set for general qubit
Markovian quantum dynamics (Theorem 3).

The most immediate open questions are related to the corresponding results
for higher dimensions.  We have seen that for $N \ge 3$, unital operations
no longer correspond to real GKS matrices.  An interesting open question is
whether there is still a {\em finite} generating set for unital Markovian
quantum dynamics.  It would also be interesting to see how the results for
general processes scale to $N \ge 3$ dimensions and how many parameters are
needed to describe the extreme points of the basis set.  This will require
investigation of the specific adjoint representation of $SU(N)$ that acts on
$(N^2-1) \times (N^2-1)$ GKS matrices.

Further open questions arise from other possible composition rules for
quantum operations.  If we lift the restriction to Markovian quantum
dynamics, it is reasonable to include composition procedures other than
linear-combination and unitary conjugation.  For instance, we might allow
direct composition of operations.  Or we might consider implementing
adaptive methods of quantum control, as suggested by Lloyd and
Viola~\cite{Viola}.  We might also combine processes probabilistically
(either by tossing coins or by performing different operations on different
parts of the sample in a bulk quantum computer), giving a convex combination
$p^i {\cal E}^i$ of quantum operations.  Such combination does not preserve
Markovity in general, but it has other attractive properties; for example,
any unital operation on a qubit can be written as a convex sum of {\em
unitary} operations (although this does not hold for higher dimensional
systems~\cite{Unital}).  The characterization of the extremal operations on
a qubit presented in~\cite{Ruskai} can be used to extend this to nonunital
operations.

Finally, we might consider a formulation of the problem which requires only
that we be able to come arbitrarily close to a given quantum
operation~\cite{Dorit}.  This would be more in line with the usual notion of
universality for unitary operations, and might prove fruitful as a way of
finding smaller basis sets for general quantum operations.

\section*{Acknowledgments}

We thank Dorit Aharonov, Daniel Lidar, and Michael Nielsen for useful
discussions.
We also thank the referee for pointing out the true origin of
Eq.~(\ref{eq:trotter}) and a possible connection between the GKS
matrix and the affine representation.
DB and JK are supported by the U.S. Army Research Office under
contract/grant number DAAG55-98-1-0371.  DL and XZ are partially
supported by the DARPA Ultrascale Program under contract
DAAG55-97-1-0341, and DL by the NSA and ARDA under ARO contract number
DAAG55-98-C-0041.  DL acknowledges partial support from IBM and NTT.

\appendix

\section{Formal properties of Markovian semigroups}
\label{app:qops}

A strongly continuous one-parameter semigroup on a complex Banach space $B$
is defined as a family ${\cal E}_t$ of bounded linear operators ${\cal
E}_t:
B \to B$ parametrized by real $t \ge 0$ which satisfy
\begin{enumerate}
\item[(a)] ${\cal E}_0={\cal I}$,
\item[(b)] ${\cal E}_s {\cal E}_t = {\cal E}_{s+t}$, and
\item[(c)] the map $(t,\rho) \to {\cal E}_t(\rho)$ from $[0,\infty)
           \times B$ to $B$ is jointly continuous.
\end{enumerate}

The generator ${\cal Z}$ of a strongly continuous one-parameter
semigroup is determined by
\be
  {\cal Z}(\rho)=\lim_{t \downarrow 0} {{\cal E}_t(\rho) - \rho \over t}
\,.
\label{eq:generator_append}
\ee
The domain of ${\cal Z}$, $\mbox{Dom}({\cal Z})$, is defined to be the
space for which the above limit exists.  $\mbox{Dom}({\cal Z})$ is a
dense linear subspace of $B$.
If $\rho \in \mbox{Dom}({\cal Z})$,
\be
  {\partial\rho(t) \over \partial t} = {\cal Z}(\rho(t))
  \quad \forall t \ge 0
\,,
\ee
and the semigroup is defined by its generator according to
\be
  {\cal E}_t = e^{{\cal Z} t} \equiv \lim_{n \to \infty}
             \left( {\cal I} - {t \over n} {\cal Z} \right)^{-n}
\,,
\label{eq:exponent_append}
\ee
the inverse being a bounded operator for sufficiently large $n$.

A one-parameter semigroup is norm continuous if and only if the generator
is bounded, in which case
\be
  {\cal E}_t = \lim_{n \to \infty} \sum_{n=0}^{\infty}
          \frac{{\cal Z}^n t^n}{n!}
\,.
\label{eq:taylor_append}
\ee

Continuous one parameter semigroups capture the Markovian and stationarity
features of the Markovian quantum dynamics of interest.  The remaining
features to be incorporated are complete positivity and trace preservation.
This leads to our definition:
\begin{quote}
  {\em A Markovian semigroup is a norm continuous one-parameter semigroup of
  completely positive, trace preserving linear maps.}
\end{quote}

\section{The affine representation and the GKS matrix}
\label{app:gks2lin}

Following the discussion in Section \ref{sec:qops}, consider a basis for
traceless operators $\{F_\alpha\}$ that is Hermitian and trace
orthonormal.  Then we can express a density matrix as $\rho = \rho_0
\, I + \sum_\alpha \rho_\alpha F_\alpha$ where $\rho_0, \rho_\alpha$
are real numbers.
Due to trace preservation, the linear representation of the generator
${\cal Z}$ of a Markovian semigroup can be reduced to an affine
map on the traceless components only:
\be
	\dot \rho_\alpha = L_{\alpha \beta} \rho_\beta + p_\alpha
\,.
\label{eq:affinez}
\ee
In Eq.~(\ref{eq:affinez}), $L_{\alpha \beta}$ are entries of an $(N^2
- 1) \times (N^2 - 1)$ matrix $L$, and $p_\alpha$ are entries of an
$(N^2 -1)$-dimensional vector.

In the qubit case, let $\{F_\alpha \} = {1 \over \sqrt{2}} \{\sigma_x,
\sigma_y, \sigma_z \}$.  Using Eq.~(\ref{eq:gks}), we obtain a
one-to-one correspondence between the GKS matrix $A$ (with entries
$a_{\alpha \beta}$) and $\{ L, p\}$ in the affine representation:
\bea
	L & = & \left( \begin{array}{ccc}
	- 2 \, (a_{22} \!+ \!a_{33}) & a_{12} + a_{21} & a_{13} + a_{31}
\\	a_{12} + a_{21} & - 2 \,(a_{11} \!+\! a_{33}) & a_{23} + a_{32}
\\	a_{13} + a_{31} & a_{23} + a_{32} & - 2 \, (a_{11} \!+\! a_{22})
	\end{array} \right)
\nonumber
\\
	& = & A + A^T - (2 \, {\rm tr} A) \, I
\,,
\\
	p & = & 4 \left({\rm Im}(a_{32}),
			{\rm Im}(a_{13}),
			{\rm Im}(a_{21}) \right)^T
\,.
\eea
As positivity of the GKS matrix is equivalent to complete positivity
of Eq.~(\ref{eq:gks}), the correspondence between $A$ and $\{L, p\}$
allows complete positivity of the affine representation to be
easily characterized.

For higher dimensions, the affine representation $\{L, p\}$ and 
the GKS matrix for a generator ${\cal Z}$ are still in one-to-one 
correspondence.  In particular, let 
$[F_\alpha, F_\beta] = i f_{\alpha \beta \gamma} F_\gamma$ and 
$\{F_\alpha,F_\beta\} = h_{\alpha \beta \gamma} F_\gamma$ where 
$f_{\alpha \beta \gamma}$ are the real structure constants and 
$h_{\alpha \beta \gamma}$ are real. 
Then the entries of $L$ and $p$ are given by    
\bea
   L_{\eta \gamma} &=& - h_{\alpha \gamma \lambda} f_{\lambda \beta \eta} 
			~{\rm Im}(a_{\alpha \beta}) 
\nonumber
\\        & & - f_{\alpha \gamma \lambda} f_{\lambda \beta \eta} 
			~{\rm Re}(a_{\alpha \beta})
\,,
\label{eq:gks2l}
\eea
\be
	p_\gamma = - 4 \sum_{\alpha < \beta} f_{\alpha \beta \gamma} 
						{\rm Im}(a_{\alpha \beta})
\,.
\label{eq:gks2p}
\ee
Conversely, given $\{L, p\}$, ${\cal Z}$ is uniquely determined. 
For a fixed basis $\{F_\alpha\}$, a unique decomposition in the form of 
Eq.~(\ref{eq:gks}) exists, so that the GKS matrix is 
uniquely determined.  In other words, the linear system  
Eqs.~(\ref{eq:gks2p})--(\ref{eq:gks2l}) can be inverted.  
Thus the $\{L, p\}$ that represent Markovian semigroups are those for which
a solution to Eqs.~(\ref{eq:gks2p})--(\ref{eq:gks2l}) exists and corresponds
to a positive semidefinite GKS matrix.  

\section{A unital process with complex GKS matrix}
\label{app:unital}

A convenient set of generators for $SU(3)$, known as the Gell-Mann matrices,
are defined as
\be
\begin{array}{rclrcl}
  \sqrt{2} \lambda_{1} &=&
  \mthree{0}{1}{0}
         {1}{0}{0}
         {0}{0}{0} &
  \sqrt{2} \lambda_{2} &=&
  \mthree{0}{-i}{0}
         {i}{0}{0}
         {0}{0}{0}
\\
  \sqrt{2} \lambda_{3} &=&
  \mthree{1}{0}{0}
         {0}{-1}{0}
         {0}{0}{0} &
  \sqrt{2} \lambda_{4} &=&
  \mthree{0}{0}{1}
         {0}{0}{0}
         {1}{0}{0}
\\
  \sqrt{2} \lambda_{5} &=&
  \mthree{0}{0}{-i}
         {0}{0}{0}
         {i}{0}{0} &
  \sqrt{2} \lambda_{6} &=&
  \mthree{0}{0}{0}
         {0}{0}{1}
         {0}{1}{0}
\\
  \sqrt{2} \lambda_{7} &=&
  \mthree{0}{0}{0}
         {0}{0}{-i}
         {0}{i}{0} &
  \sqrt{6} \lambda_{8} &=&
  \mthree{1}{0}{0}
         {0}{1}{0}
         {0}{0}{-2}
\,.
\end{array}
\ee
The only nonvanishing structure constants are $f_{321} = -\sqrt{2}$,
$f_{651} = 1/\sqrt{2}$, $f_{642} = f_{741} = f_{752} = -1/\sqrt{2}$,
$f_{854} = f_{876} = -\sqrt{3}/\sqrt{2}$ together with cyclic permutations
of the indices.

The following process defined over the Gell-Mann matrices is unital,
although $A$ is complex:
\be
  A = \frac{1}{2}
      \left( \matrix{0 & 0 & 0 & 0  &  0 & 0 & 0 & 0 \cr
                     0 & 0 & 0 & 0  &  0 & 0 & 0 & 0 \cr
                     0 & 0 & 0 & 0  &  0 & 0 & 0 & 0 \cr
                     0 & 0 & 0 & 1  &  0 & 0 & i & 0 \cr
                     0 & 0 & 0 & 0  &  1 & i & 0 & 0 \cr
                     0 & 0 & 0 & 0  & -i & 1 & 0 & 0 \cr
                     0 & 0 & 0 & -i &  0 & 0 & 1 & 0 \cr
                     0 & 0 & 0 & 0  &  0 & 0 & 0 & 0} \right)
\,.
\ee
The eigenvalues are $2$ and $0$ with degeneracies $2$ and $6$, so the
process is well defined.  To show that it is unital, we need to
show $\sum_{\alpha < \beta} a_{\alpha \beta} f_{\alpha\beta\gamma} = 0$
$\forall \gamma$.  First, $a_{\alpha\beta} = 0$ for all $\alpha < \beta$
except for $a_{47} = a_{56} = i$.  The criterion reduces to $f_{47\gamma} +
f_{56\gamma} = 0$ $\forall \gamma$.  But $f_{47\gamma} = f_{56\gamma} = 0$
$\forall \gamma \ne 1$, and $f_{471} = 1/2$, $f_{561} = -1/2$.  Therefore,
the process is unital.



\begin{thebibliography}{20}
\bibitem{Dorit}
  This was suggested to us by Dorit Aharonov.
\end{thebibliography}
\end{document}